\documentclass[aps,floatfix,preprint,nofootinbib]{revtex4}
\usepackage{graphics}
\usepackage{graphicx}
\usepackage{amssymb}
\usepackage{amsmath}
\usepackage{verbatim}
\usepackage{pstricks}
\newcommand{\be}{\begin{equation}}
\newcommand{\bea}{\begin{eqnarray}}
\newcommand{\beq}[1]{\begin{equation}\label{#1}}

\newcommand{\ee}{\end{equation}}
\newcommand{\eea}{\end{eqnarray}}
\newcommand{\eeq}{\end{equation}}
\newcommand{\lsim}{\!\mathrel{\hbox{\rlap{\lower.55ex \hbox{$\sim$}} \kern-.34em \raise.4ex \hbox{$<$}}}}
\newcommand{\gsim}{\!\mathrel{\hbox{\rlap{\lower.55ex \hbox{$\sim$}} \kern-.34em \raise.4ex \hbox{$>$}}}}

\newcommand{\mr}[1]{\mathrm{#1} }

\begin{document}

\setlength{\baselineskip}{0.22in}

\begin{flushright}MCTP-09-41 \\
\end{flushright}
\vspace{0.2cm}

\title{Supersymmetric Baryogenesis from Exotic Quark Decays}

\author{Timothy Cohen, Daniel J. Phalen and Aaron Pierce}
\vspace{0.2cm}
\affiliation{Michigan Center for Theoretical Physics (MCTP) \\
Department of Physics, University of Michigan, Ann Arbor, MI
48109}

\date{\today}

\begin{abstract}
In a simple extension of the minimal supersymmetric standard model, out-of-equilibrium decays of TeV scale exotic vector-like squarks may generate the baryon asymmetry of the universe.  Baryon number and CP violation are present in the superpotential, so this mechanism does not rely on CP violation in supersymmetry breaking parameters.  We discuss phenomenological constraints on the model as well as potential signals for the Large Hadron Collider and electronic dipole moment experiments.  A variation on the TeV scale model allows the exotic squarks to be the messengers of gauge mediated supersymmetry breaking.

\end{abstract}

\maketitle

\section{Introduction}
There is a large literature on TeV-scale exotic quarks and squarks (see \cite{Kang:2007ib} and the references therein).  In addition to being of immediate interest for collider searches, they are present in many supersymmetric (SUSY) grand unified theories (GUT) and in string based constructions. The messenger sector of gauge mediated models of supersymmetry breaking (GMSB) also contains such matter \cite{Dine:1982zb, AlvarezGaume:1981wy, Dine:1993yw}.  In this paper, we investigate the possibility that decays of vector-like quarks might have sourced the baryon asymmetry of the universe (BAU).  In contrast  to many models of high-scale baryogenesis, this model presents the tantalizing possibility of probing the physics of baryon number generation at future colliders.

To the Minimal Supersymmetric Standard Model (MSSM), we add exotic vector-like quark supermultiplets with renormalizable baryon number violating interactions in the superpotential.  The decays of the lightest exotic particle (LXP) source the BAU.  For concreteness, we focus on scenarios where the LXP is a squark, rather than its fermionic partner.  While it is possible to generate the BAU with generic TeV-scale masses for the exotic squarks,   there are also interesting regions of parameter space where the BAU is achieved via a resonant enhancement which requires highly degenerate exotic squark masses.

The CP violation responsible for the BAU is independent of supersymmetry breaking.  So far the only non-zero phase observed in nature is the Cabibbo-Kobayashi-Maskawa (CKM) phase which resides in the superpotential.  Constraints on new CP violating processes induced by SUSY breaking are stringent.  Perhaps nature has only chosen to have large CP violating phases in supersymmetric terms\footnote{We thank M.~Luty for discussions of this point. }. If nature chooses this path, models that rely on SUSY breaking phases cannot generate the baryon asymmetry, and a model such as the one presented here would be required.

Previous work has noted the possibility that out-of-equilibrium, baryon number violating, superpartner decays might generate the BAU \cite{Dimopoulos:1987rk,Adhikari:1996mc, Huber:2005iz, Cline:1990bw, Mollerach:1991mu}.  In particular, Dimopoulos and Hall \cite{Dimopoulos:1987rk} used the baryon number violating operator $u^{c} \, d^{c} \, d^{c}$ to produce the BAU from the decay of MSSM squarks.  In contrast to our approach, CP violation in that model derives from SUSY breaking terms.  

There exist a variety of phenomenological constraints and future tests of our model.
While our approach is largely insulated from the phenomenological difficulties associated with EDMs, it predicts values which could be seen in future experiments.   Unitarity of the CKM matrix and $D^0-\overline{D}^0$ mixing yield important restrictions.  For favorable parameters, one could perhaps observe baryon number violation directly at the LHC.  Cosmological considerations require the reheat temperature to be at most $\mathcal{O}$(10 GeV) for the TeV scale exotics, high enough to allow for a thermal weak scale dark matter candidate \cite{Lee:1977ua}, but low enough to avoid the gravitino \cite{Weinberg:1982zq} and moduli \cite{Coughlan:1983ci, Banks:1993en, deCarlos:1993jw} problems.

In the next section we describe the details of the model.  In Sec. \ref{sec:BaryoFromDecays} we calculate the asymmetry from the LXP squark decay.  In Sec. \ref{sec:Cosmology} we outline the cosmology and related constraints.  In Sec. \ref{sec:gaugeMediation} we discuss the variation when the exotics are the messengers of gauge mediation.  In Sec. \ref{sec:Pheno} we discuss the low energy observables and collider signatures.  The appendices discuss model building challenges for degenerate exotic squarks and provide explicit estimates of the cosmological rates for Sec. \ref{sec:Cosmology}.

\section{The Model}\label{sec:theModel}
The relevant matter is the three generations of colored MSSM chiral superfields ($u^c_i$, $d^c_i$, $q_i$), $i=1\ldots 3$, supplemented by $N$ families of exotic vector-like quark superfields ($D_i$,\,$\overline{D}_i$), $i=1\ldots N$. We concentrate on the model with $N=2$, which is the simplest case where this mechanism is viable.  There is an approximate $\mathbb{Z}_2$ ``exotic-parity" under which the $D$ and $\overline{D}$ are odd while all other superfields are even.  If this parity were exact, the LXP would be stable.  The decays of the LXP generate the BAU. The superpotential is
\be
\mathcal{W} = \mathcal{W}_\mr{MSSM} + \mathcal{W}_\mr{Exotic},
\ee
with
\be
\mathcal{W}_\mr{Exotic} = g'_{ijk}\,u^c_i\,D_j\,D_k +
(\mu'_R)_{ij}\,d^c_i\,\overline{D}_j + \left(\frac{(\mu'_L)_{ij}}{v_d}\right)H_d\,q_i\,D_j + M_{ij}\,D_i\,\overline{D}_j,
\label{eq:W_Exotic}
\ee
where $H_d$ is the MSSM Higgs which couples to down-type fields; $v_d \equiv \langle H_d \rangle$; $g'$ is a $B$-violating coupling between the MSSM and exotic sectors; $\mu'_L$ and $\mu'_R$ are (small) exotic-parity violating couplings, and $M$ is the mass matrix for the exotics.  This superpotential is in a basis where Standard Model Yukawa couplings have been diagonalized, and there is no mixing between the MSSM and exotic (s)quarks at gauge boson/gaugino vertices.  In estimates below we use a common exotic-parity violating coupling, $\mu' \equiv \mu'_L = \mu'_R$.  Dependence on either $\mu'_L$ or $\mu'_R$ follows from Eqs.~(\ref{eq:DdZCoupling}), (\ref{eq:uDdCoupling}), and (\ref{eq:uddCoupling}).  Since $\mu'\ll M$ for all viable models, we work to lowest order in $\mu'$.  We have omitted similar couplings of the form $u^c\,d^c\,d^c$ and $u^c\,d^c\,D$.  Such $B$-violating couplings are present after a rotation to eliminate the $\mu'$ terms.  The assumption that such couplings are negligible prior to rotation motivates the hierarchy in Eqs. (\ref{eq:uDdCoupling}) and (\ref{eq:uddCoupling})\footnote{A $\mathbb{Z}_4$ extension of R-parity (which would also be broken by $\mu'$) can be constructed to impose the vanishing of the $u^c\,d^c\,d^c$ and $u^c\,d^c\,D$ couplings.}.

The superpotential in Eq.~(\ref{eq:W_Exotic}) satisfies two of the Sakharov conditions \cite{Sakharov:1967dj}: $B$ and CP violation. In the presence of both $g'$ and $\mu'$, it is impossible to consistently assign baryon number, and there are physical CP violating phases for $N \geq 2$.  Using field redefinitions of the $u^c_i$ which leave the mass matrix diagonal, one can always make $g'_{ijk}$ real for $N=2$.  Phases remain in the $\mu'$ matrices. As we discuss in Sec. \ref{sec:Cosmology},  the out-of-equilibrium condition is dictated by the cosmology:  we imagine that the late decay of a modulus reheats the universe and (over)populates the LXP.

We demonstrate the diverse phenomenology of the $(g',\,\mu')$ parameter space by presenting three scenarios which (see Table  \ref{tab:benchmarkParams})
\begin{enumerate}
\item[i.] have generic TeV scale masses for the exotic squarks,
\item[ii.] maximize the reheat temperature of the universe, thereby requiring degenerate TeV scale squarks,
\item[iii.] identify the exotics with the messengers of GMSB.
\end{enumerate}
 Also shown in Table \ref{tab:benchmarkParams} are the exotic squark mass and the splitting between the two lightest squarks.  For the degenerate $\tilde{D}$ mass benchmarks (ii. and iii. above), all splittings are at the sub-percent level, which leads to a resonant enhancement of the BAU.  

We assume no CP violation in the SUSY breaking sector, consistent with our philosophy that all CP violation comes from superpotential couplings.  Additionally, this both simplifies the analysis and highlights differences between our model and that of \cite{Dimopoulos:1987rk}, where the phase arises from soft-terms.  While we are agnostic about the origin of the $M$ and $\mu^{\prime}$ terms, for the large splittings and high reheat parameters in Table \ref{tab:benchmarkParams}, the Giudice-Masiero mechanism \cite{Giudice:1988yz} might be responsible for their origin (perhaps with a loop-factor generating the hierarchy between them).

\begin{table}[h]
\begin{center}
\begin{tabular}{||c|c|c|c|c|c||}

\hline
\hline
\multicolumn{6}{|c|}{Large $\tilde{D}$ mass splittings} \\
\hline
scenario           & $g'$ & $\frac{\tilde{M}}{(\mr{GeV})}$ & $\frac{\mu'}{(\mr{GeV})}$ & $\frac{\Delta \tilde{M}^2}{(\mr{GeV})^2}$ & $\epsilon$ \\
\hline
large splittings   &  0.4   & $500$  & $4$ &  $(100\,\mu')^2$  & $2\times 10^{-6}$\\
\hline
\hline
\multicolumn{6}{|c|}{Degenerate $\tilde{D}$ masses} \\
\hline
scenario           & $g'$ & $\frac{\tilde{M}}{(\mr{GeV})}$ & $\frac{\mu'}{(\mr{GeV})}$ & $\frac{\Delta \tilde{M}^2}{(\mr{GeV})^2}$ & $\epsilon$ \\
\hline
high $T_\mr{RH}$  &  0.005    &  1000  & 2  & $(1.3\,\mu')^2$ & $8\times 10^{-6}$\\
gauge mediation   &   0.01  &  $10^6$  & $1$ &  $(1.0\,\mu')^2$  & $6\times 10^{-5}$\\
\hline
\hline
\end{tabular}
\caption{Benchmark parameters for three different scenarios.  The first has generic values for the lightest two $D$ squarks.  The last two rely on degenerate squarks to give a resonant enhancement of $\epsilon$:  a high reheat temperature scenario, and a case where the exotics are identified with the messengers of gauge mediation.  In the last two cases, we assume all the $g'_{ijk}$ and $\mu'_{ij}$ are approximately independent of family.   For the first (large splittings) benchmark we require hierarchies in these values to avoid the phenomenological bounds alluded to in Sec. \ref{sec:textures}.}
\label{tab:benchmarkParams}
\end{center}
\end{table}

A SUSY breaking term $b_M\tilde{D}\,\tilde{\overline{D}}$ splits the squark masses, resulting in light and heavy mass eigenstates, $(\tilde{D}_{\ell})_i$ and $(\tilde{D}_h)_i$, where $i=1,2$ for $N=2$.  Unless the non-holomorphic contributions to the squark mass are large, there is an exotic squark lighter then the exotic quarks.  Should the non-holomorphic SUSY-breaking make the LXP a fermion, the generation of the asymmetry proceeds in a nearly identical fashion through the decay of the exotic quarks.  For unity of discussion, we will assume a squark LXP for all benchmarks.

The mass difference $\Delta \tilde{M}^{2} \equiv (\tilde{M}_{\ell})_2^2-(\tilde{M}_{\ell})_1^2$ between the two lightest exotics has a large impact on the size of the BAU generated. A near degeneracy yields a resonant enhancement of the baryon asymmetry \cite{Flanz:1996fb, Pilaftsis:1997jf, Pilaftsis:2003gt}.  We discuss this possibility in detail in Sec. \ref{sec:BaryoFromDecays}.  When required, to achieve a degenerate spectrum we assume that some symmetry enforces degenerate values of $M$ for the two families of exotics.  The symmetry can be broken by the $\mu'$ terms, so we assume an $\mathcal{O}(1)$ generation dependence in $\mu'_{ij}$.  This implies that the lightest (heaviest) two exotic squarks, $(\tilde{D}_{\ell})_i$ ($(\tilde{D}_h)_i$) are degenerate in mass up to $\mu'^2$ corrections.  This motivates the parametrization of $\Delta \tilde{M}^2$ in Table \ref{tab:benchmarkParams}.  Since each $\tilde{D}_{\ell}$ is an equal admixture of $\tilde{D}$ and $\tilde{\overline{D}}^*$, there are additional factors of $1/\sqrt{2}$ introduced into the interactions relevant for the exotic decays.  For clarity, we work with the ``helicity" squark eigenstates, $\tilde{D}$ and $\tilde{\overline{D}}^*$.  We refer to both $(\tilde{D}_{\ell})$s as LXPs since both can potentially contribute to the BAU.

The degeneracy between the two LXP states can be broken by either off-diagonal elements in the SUSY and SUSY-breaking masses or by radiative corrections.  Since degeneracy is important for benchmark points ii. and iii., it is important that these terms can be made small.   The absence of these terms can be understood in terms of the same (almost) conserved family symmetry mentioned above.  Some relevant model building issues are discussed in Appendix \ref{sec:SmallSplit}.  To keep expressions simple we will often use $M$ for both the SUSY mass parameter and the mass of the LXP when estimating various processes.

\subsection{Interactions Induced by Diagonalization}\label{sec:InteractionsInSCKMBasis}

Mass diagonalization mixes $d_L$ with $\overline{D}$ and $d_R$ with $D$.  We do not introduce distinct notation for gauge and mass eigenstate fields. In what follows, $d_{L}$, for example, is the mass eigenstate state with the largest overlap with the $d_L$ from above.  Because $d_L$ and $\overline{D}$ have different electroweak charges, the rotation to the mass eigenbasis induces the following off-diagonal couplings to the $Z^0$ boson
\be\label{eq:DdZCoupling}
\left(\frac{1}{2}\,\frac{g_w}{c_w}\,\frac{\mu_L'}{M}\right)(d_L)^{\dagger}\overline{\sigma}^{\mu}(\overline{D})\,Z^0_{\mu} + \mr{h.c.}
\ee
where $g_w$ is the $SU(2)$ coupling constant and $c_w \equiv  \cos \theta_w$.  There are also MSSM-exotic couplings with the $W^{\pm}$, as well as the supersymmetric analogues of both of these interactions.

There is one other class of interactions important for this study.  After the $d_R-D$ rotation, the following couplings appear:
\bea
&g'&\left(\frac{\mu'_R}{M}\right)\,u^c\,d^c\,D  \label{eq:uDdCoupling}\\
&g'&\left(\frac{\mu'_R}{M} \right)^2 u^c\,d^c\,d^c.\label{eq:uddCoupling}
\eea
The generation of the BAU relies on the interactions in Eqs. (\ref{eq:DdZCoupling}) and (\ref{eq:uDdCoupling}).  If either $\mu'_{L}$ or $\mu'_{R}$ were to vanish, then the BAU generated would be suppressed by powers of $m_{d} /M$, where $m_{d}$ is a Standard Model down-type quark mass.  The interaction of Eq.~(\ref{eq:uddCoupling}) is a standard MSSM R-parity violating coupling.   It is small because $\mu'/M \ll 1$.  For the purposes of this model, the dominant effect of these operators is to cause the LSP to decay, which is relevant both for collider signatures and for the cosmology.

We assume off-diagonal gauge interactions between the exotic and MSSM sectors induced by soft-terms are sub-dominant for the purposes of the calculation of the BAU.  We assume this both for simplicity and because we wish to emphasize that this mechanism can occur independent of SUSY breaking.

\subsection{Textures in $g'$ and $\mu'$}\label{sec:textures}
One possibility is that all $g'_{ijk}$ (and all the $\mu'_{ij}$) are comparable. Under this assumption, phenomenological constraints make it difficult to realize the BAU without a resonant enhancement due to degenerate squarks (more on this in Sec. \ref{sec:BaryoFromDecays}).  However, these constraints (e.g. $D^0-\overline{D}^0$ mixing) only pertain to specific families, and can be avoided if hierarchies exist in these couplings.  We assume such textures apply for the ``large splittings scenario'' of Table \ref{tab:benchmarkParams}.  In this case, the values shown in Table \ref{tab:benchmarkParams} are the biggest entries in the $g'$ and $\mu'$ matrices -- they lead to the dominant contribution to the BAU.

There is another potential motivation for textures in $\mu'$.  The lightest eigenvalues of the full quark mass matrix, which would correspond to the MSSM down-type quarks, are given by a see-saw:
\be
m_d^\mr{physical} = m_d - \frac{\mu'_L\,\mu'_R}{M}.
\label{eq:finetuning}
\ee
Without textures in the $\mu'$ matrices, $M = 500$ GeV implies that $\mu' \lesssim 2$ GeV to avoid fine-tuning between the two contributions to the down quark mass.  Hence, for the large splittings benchmark we assume there is a texture which eliminates this tuning for the down quark.  This can be done without eliminating all of the CP violating phases.  For the benchmarks with degenerate LXPs $\mu'$ is already small enough to avoid fine-tuning.

\section{Baryogenesis from Exotic Squark Decays}\label{sec:BaryoFromDecays}

Our goal is to reproduce the BAU, accurately measured by the WMAP5 \cite{Dunkley:2008ie} data to be
\be\label{eq:etaWMAP}
\eta\equiv \frac{n_B-n_{\overline{B}}}{n_{\gamma}} = 6.225\pm0.170\times 10^{-10},
\ee
where $n_B$ ($n_{\overline{B}}$) is the number density of baryons (anti-baryons) and $n_{\gamma}$ is the number density of photons in the universe.

A calculation of the baryon asymmetry requires two ingredients: a knowledge of the co-moving number density of exotics $n_{D}/s$ when they decay out-of-equilibrium and the baryon asymmetry created in each decay ($\epsilon$). A necessary condition for the squarks to be out-of-equilibrium is that the annihilations are no longer effective. When annihilations eventually do freeze-out, the resulting $n_{D}/s$ produces an insufficient $\eta$.  Hence, the squarks must be populated by some non-thermal source.   For our benchmarks, non-thermal decays of a heavy field (see Sec. \ref{sec:Cosmology}) generate an $n_D/s$ in the range $10^{-4}-10^{-6}$, necessitating an $\epsilon \sim 10^{-4}-10^{-6}$ to reproduce the measured $\eta$.  In this section, we discuss the calculation of $\epsilon$ and postpone a detailed discussion of  $n_{D}/s$ to the next section.

The LXP decays yield a net baryon number per decay
\be
\epsilon \equiv \sum_{i=1}^n\frac{\Gamma((\tilde{D}_{\ell})_i\rightarrow u+d)-\Gamma((\tilde{D}_{\ell}^*)_i\rightarrow
u^{\dagger}+d^{\dagger})}{\Gamma_\mr{total}((\tilde{D}_{\ell})_i)},
\ee
where $n$ is the number of squarks that make non-trivial contributions.  Since $B$-violating decays to the MSSM states are suppressed by $\mu'^2$, exotic states have small partial widths for these processes.  However, exotic-parity ensures the \emph{total} width of the lightest exotics are also suppressed by $\mu'^2$.  This allows the $B$-violating decays for LXP states to compete with the total width, yielding an $\epsilon$ of appreciable size.

\begin{figure}[ht]
\begin{center}
\includegraphics[scale=1]{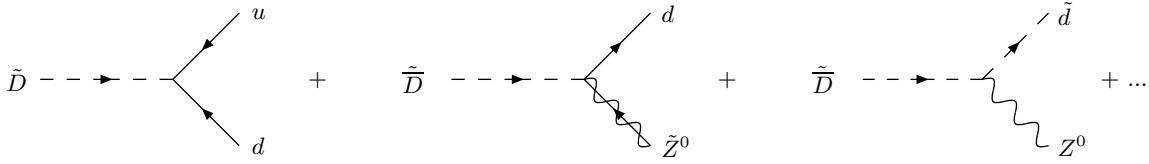}
\end{center}
\caption{Representative contributions to $\Gamma^\mr{total}_{\ell}$ for the lightest $D$ squarks.  For the purposes of our estimates we assume the third process is kinematically allowed for all three families of $\tilde{d}$.}
\label{fig:GammaTotalD}
\end{figure}

First we estimate $\epsilon$ neglecting possible complications due to resonance.  We consider the decays in Fig.~\ref{fig:GammaTotalD}\footnote{There are a number of sources of order one uncertainty in our estimates.  Since we are not interested in a detailed exploration of the parameter space, but rather a demonstration of the viability of this approach, we will not worry about errors of this size -- a small change in the input parameters can compensate.  In this spirit, we display (and compute) representative contributions to $\epsilon$, but do not make an exhaustive calculation of all diagrams.  For example, we neglect processes involving the $W^{\pm}$ and their superpartners and LXP decays to $\tilde{u}+\tilde{d}$.  We also assume that $\tilde{Z}^0$ is a mass eigenstate and neglect corrections due to neutralino mixing.}.  Including the sum over all possible quark final states gives a width:
\be\label{eq:TotalWidth}
\Gamma^\mr{total}_{\ell} \approx
\frac{9\,g'^2+6\,(g_w/(2\,c_w))^2}{16\,\pi}\,\frac{1}{2}\,\left(\frac{\mu'}{M}\right)^2\,\tilde{M}.
\ee

\begin{figure}[ht]
  \begin{center}
\includegraphics[scale=1]{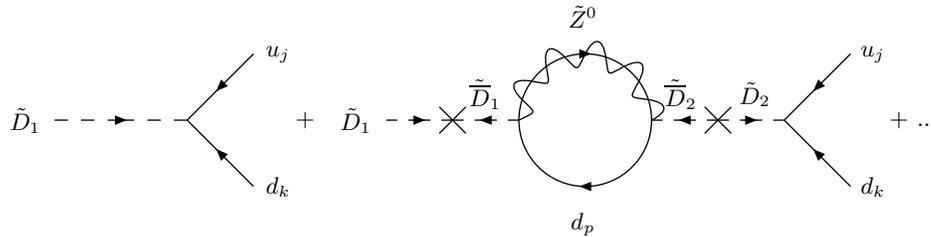}
\end{center}
\caption{Representative diagrams that interfere to give a net baryon number.  The squark mass insertions illustrate which ``helicity" component of the $(\tilde{D}_{\ell})_i$ is contributing to the amplitude.}
\label{fig:epsilon}
\end{figure}

A representative contribution to $\epsilon$ is given in Fig. \ref{fig:epsilon}. We assume $\tilde{D}$ can decay to (three families of) $\tilde{d}$ and $Z^0$ as well, so there will also be intermediate loops involving these particles, which we include in our estimate of $\epsilon$.  We estimate
\be
\epsilon_\mr{no-resonance} \approx \frac{3}{16\,\pi}\,\frac {\,g'^2\,\left(\frac{g_w}{c_w}\right)^2}{9\,g'^2+\frac{3}{2}\,\left(\frac{g_w}{c_w}\right)^2}\left(\frac{\mu'^2}{\Delta \tilde{M}^2}\right),
\ee
where we have included contributions from both of the light $\tilde{D}$s, 
and defined $\Delta \tilde{M}^2 \equiv (\tilde{M}_{\ell})_2^2-(\tilde{M}_{\ell})_1^2$.  In this estimate, we assume all $\mu'$ and $g'$'s are the same order, and assume a single non-zero (maximal) phase.  Motivated by phenomenological constraints discussed below, we take $\mu'/M \sim 10^{-2}$, $g'\sim 10^{-2}$ which gives
\be
\epsilon_\mr{no-resonance} \sim 10^{-10}\,\left(\frac{g'}{10^{-2}}\right)^2\,\left(\frac{\mu'/M}{10^{-2}}\right)^2\left(\frac{\tilde{M}^2}{\Delta \tilde{M}^2}\right).
\ee
Since we are trying to achieve an $\epsilon\, \gsim 10^{-6}$, it is clear why a resonant enhancement is necessary for much of the parameter space.  The need for resonance can only be avoided for larger $g'$, which requires a texture to avoid flavor changing neutral current constraints.

\subsection{Full Calculation of $\epsilon$ Including Resonance Effect}\label{sec:FullCalcEpsilon}
We now calculate $\epsilon$ accounting for the possibility of highly degenerate $(\tilde{D}_{\ell})$s following \cite{Pilaftsis:1997jf, Pilaftsis:2003gt} .  Using the couplings of Eqs. (\ref{eq:DdZCoupling}) and (\ref{eq:uDdCoupling}) gives

\bea
\epsilon &\approx& \frac{\left(\frac{g_w}{2\,c_w}\right)^2\mr{Im}[(g')^*_{j21}\,(g')_{j12}\,(\mu'_R)^*_{k2}\,(\mu'_R)_{k1}\,(\mu'_L)^*_{p2}\,(\mu'_L)_{p1}]}{(16\,\pi)^2\,\frac{\Gamma^\mr{total}_1}{\tilde{M}_1}\,\frac{\Gamma^\mr{total}_2}{\tilde{M}_2}}\,\frac{1}{M_1^2\,M_2^2}\times\nonumber\\
&& \,\,\left\{\frac{(\tilde{M}_1^2-\tilde{M}_2^2)\,\Gamma^\mr{total}_2\,\frac{\tilde{M}_1^2}{\tilde{M}_2}}{(\tilde{M}_1^2-\tilde{M}_2^2)^2+\left(\Gamma^\mr{total}_2\frac{\tilde{M}_1^2}{\tilde{M}_2}\right)^2} -  (1 \leftrightarrow 2) \right\}, \label{eq:epsilonBeforeApprox}
\eea
where $\tilde{M}_i \equiv (\tilde{M}_{\ell})_i$, $M_i \equiv M_{ii}$ and a sum over $j,\,k,\,p = 1\ldots 3$ is implied.  We assume the physical phase in one of the superpotential couplings $(g'\,\mu'_R)_{ijk}(u^c_i\,d^c_j\,D_k)$ equals $\pi/2$.  The value of $\epsilon$ for each benchmark is given in Table \ref{tab:benchmarkParams}.

We parametrize the mass splitting by $(x\,\mu')^2 \equiv ((\tilde{M}_{\ell})_2^2-(\tilde{M}_{\ell})_1^2) \ll \tilde{M}^2$. Then the $\mu'$ dependence essentially cancels.  This illustrates how the resonance effect can compensate for small values of $\mu'$.

\begin{figure}[t]
\includegraphics[scale=1]{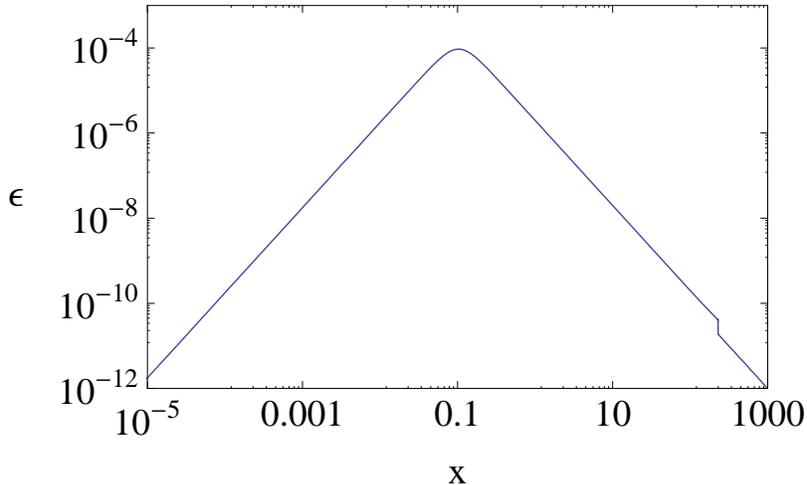}
\caption{Plot of $\epsilon$ as a function of the $\tilde{D}$ mass splitting parameter $x$ for the high reheat benchmark (see Table \ref{tab:benchmarkParams}).
Note that $\epsilon_\mr{max}\approx 9\times 10^{-5}$ for $x=0.1$.  We have also taken into account the change in $\Gamma^\mr{total}_2$ at $x=230$ when new on-shell decay channels open up (see text for discussion).}
\label{fig:PlotOfEpsilon}
\end{figure}

We have plotted $\epsilon$ in Fig. \ref{fig:PlotOfEpsilon} for the high reheat parameters.  
Note that $\epsilon\rightarrow 0$ as $x \rightarrow 0$: in the limit that the squark masses are degenerate, $(\tilde{D}_{\ell})_1$ is indistinguishable from $(\tilde{D}_{\ell})_2$, and it is not meaningful to interfere the two.  It reaches a maximum $\epsilon_\mr{max}\approx 9\times 10^{-5}$ for $x=0.1$.  
This plot has been made using a relatively small value of $g' = 5 \times 10^{-3}$.  Larger values of $g'$ would allow even larger values of  $\epsilon_\mr{max}$.  Therefore, if splittings are of order $\mu'^2$, i.e. $x\sim\mathcal{O}(1)$, we can achieve $\epsilon \gg 10^{-10}$, as required to generate the baryon asymmetry.  One also has to keep track of which $\tilde{D}_{\ell}$ states contribute to $\epsilon$.  In Eq. (\ref{eq:epsilonBeforeApprox}) we have assumed that both $\Gamma^\mr{total}_1$ and $\Gamma^\mr{total}_2$ are approximately given by $\Gamma^\mr{total}_{\ell}$ in Eq. (\ref{eq:TotalWidth}) which implies both $(\tilde{D}_{\ell})_1$ and $(\tilde{D}_{\ell})_2$ decays can contribute.  However, this is only valid when the following decays are kinematically forbidden:  $(\tilde{D}_{\ell})_2 \rightarrow D_1^{\dagger} +u$ and $(\tilde{D}_{\ell})_2 \rightarrow (\tilde{D}_{\ell})_1^{\ast}+\tilde{\chi}^0 +u$, where $\tilde{\chi}^0$ is the lightest neutralino.  We assume the first decay is forbidden, as determined by the relative size of $\tilde{M}$ and $M$.  For the high reheat parameters the second decay opens up at $x\approx 230$ for $m_{\tilde{\chi}^0} = 100$ GeV.  Once this decay channel opens, the contribution to $\epsilon$ from $(\tilde{D}_{\ell})_2$ becomes negligible and the $(1 \leftrightarrow 2)$ portion of the curly braces in Eq. (\ref{eq:epsilonBeforeApprox}) drops out.  This accounts for the apparent discontinuity in Fig. \ref{fig:PlotOfEpsilon} at $x = 230$; this is the point when the $(\tilde{D}_{\ell})_2$ contributions no longer contribute.  Then $\epsilon \sim 1/x^2$ as $x\rightarrow \infty$.

\section{Cosmology}\label{sec:Cosmology}
If the universe followed a thermal history, for $M \lesssim 10^{13}$ GeV, annihilations would keep the squarks in equilibrium until they became non-relativistic.  Assuming they lived long enough to satisfy the out-of-equilibrium condition, the squarks ultimately would have frozen-out with $n_D/s \ll \eta$, so their decays could not have generated the BAU.  Therefore, there must be a non-thermal source for the LXPs.  The relevant cosmology begins with a universe dominated by a long-lived state, $\phi$, which could be the inflaton or some other modulus.  Its decays populate the LXPs.

\subsection{Asymmetry Generated in Decay}\label{sec:EpochOfInflatonDecay}
Given a branching ratio (BR) for $\phi$ decaying into the exotic squarks, we approximate the co-moving abundance of the LXPs as \cite{Kolb:1990vq}:
\be \label{eq:YD}
Y_{D} \equiv \frac{n_{D}}{s}\approx\mr{BR}\left(\frac{T_\mr{RH}}{m_{\phi}}\right),
\ee
where $n_D$ is the number density of LXPs; $s$ is the entropy density of the universe; $T_\mr{RH}$ is the temperature of the universe generated by the $\phi$ decays, and $m_{\phi}$ is the mass of the $\phi$ field.
Once the exotics are produced, they must decay before they annihilate back to equilibrium, i.e.,
\be\label{eq:DecayRatevsAnnRate}
\Gamma_\mr{decay} > \Gamma_\mr{ann}(T_\mr{RH}),
\ee
where the annihilation rate at the reheat temperature $\Gamma_\mr{ann}(T_\mr{RH})$ is
\be \label{eq:GammaAnn}
\Gamma_\mr{ann}(T_\mr{RH}) \equiv Y_{D}\,s\,\langle \sigma_\mr{ann}\,v \rangle =
\mr{BR}\,g_*\,\frac{2\,\pi^2}{45}\,\frac{T_\mr{RH}^4}{m_{\phi}}\,\langle \sigma_\mr{ann}\,v \rangle.
\ee
Here, $g_*$ is the number of relativistic degrees of freedom.  For this model, one important annihilation process is $\tilde{D}+\tilde{D}^{*} \rightarrow g + g$ where $g$ is a gluon.  We estimate the thermally averaged annihilation cross section as
\be\label{eq:sigmavAnn}
\langle \sigma_\mr{ann}\,v\rangle \approx
\left(\frac{g_s^2}{4\,\pi}\right)^2\,\frac{1}{M^2},
\ee
where $g_s$ is the strong force coupling constant.

We also check
\be\label{eq:DecayRateVsHubble}
\Gamma_\mr{decay} > H(T_\mr{RH}),
\ee
where $H(T_\mr{RH})$ is the Hubble rate evaluated at the reheat temperature.  This means the exotic squarks decay
``instantaneously," i.e. when the temperature is still $T_\mr{RH}$.

Then the generated baryon asymmetry is given by
\be
\eta \equiv \frac{n_B-n_{\overline{B}}}{n_{\gamma}} = 7.04\,\epsilon \left(\frac{n_{D}}{s}\right) =
7.04\,\epsilon\,\mr{BR}\,\left(\frac{T_\mr{RH}}{m_{\phi}}\right),
\ee
where $\epsilon$ parametrizes the amount of baryon number violation produced by each exotic decay
(see Sec. \ref{sec:BaryoFromDecays} for the calculation of $\epsilon$) and the factor of 7.04 is from the ratio $(s/n_{\gamma})_\mr{today}$.  The last task is to make sure that this BAU survives to the present day.

\subsection{Washout Processes}\label{sec:Washout}
There are processes which can washout the BAU.  Requiring them to be ineffective  constrains the maximum reheat temperature.
Examples of the most dangerous of these baryon number violating processes are 
\bea
u + d &\rightarrow& \tilde{D}, \label{eq:inverseDecay} \\
u + \tilde{Z}^0 &\leftrightarrow& d^{\dagger}+d^{\dagger}, \label{eq:washout1}\\
u + g&\leftrightarrow& D+\tilde{D}, \label{eq:washout2}
\eea
where we assume that $\tilde{Z}^0$ is the LSP.  The dominant washout process involving only MSSM states (Eq. (\ref{eq:washout1})) will always include the LSP, since the LSP suffers the least Boltzmann suppression.  The process in Eq. (\ref{eq:inverseDecay}) is known as inverse decay (ID) and is proportional to Exp$(-M/T_\mr{RH})$.  The rate for the process in Eq. (\ref{eq:washout1}) receives the Boltzmann suppression for the one heavy initial state (Exp$(-\tilde{m}_{Z^0}/T_\mr{RH})$) when the process goes from left to right.  Due to mixing effects (see Eq. (\ref{eq:uddCoupling})) the cross section gets additional suppression ($\sigma \sim (\mu'/M)^4$).  Since the final states are effectively massless at $T_\mr{RH}$, we will refer to these processes as ``light."  For the GMSB benchmark, $T_\mr{RH} \gg m_\mr{LSP}$, so there will be no Boltzmann suppression.  In this case the rate will be negligible due to $(\mu'/M)^4$ suppression (see Appendix \ref{sec:extimateWashoutRates} for details).  The thermally averaged cross sections for diagrams like those in Eq. (\ref{eq:washout2}) do not depend on $\mu'$ but do suffer Boltzmann suppression for $T_\mr{RH} < M$ due to the heavy final states when the process goes from left to right.  Therefore, we will refer to these processes as ``heavy."

\begin{figure}[ht]
\begin{center}
\includegraphics[scale=1]{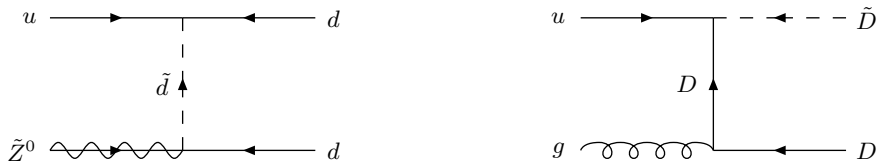}
\end{center}

\caption{Example diagrams for washout processes.}
\label{fig:WashoutProcesses}
\end{figure}

\begin{figure}[t]
\centering
\begin{tabular}{cc}
\includegraphics[scale=0.7]{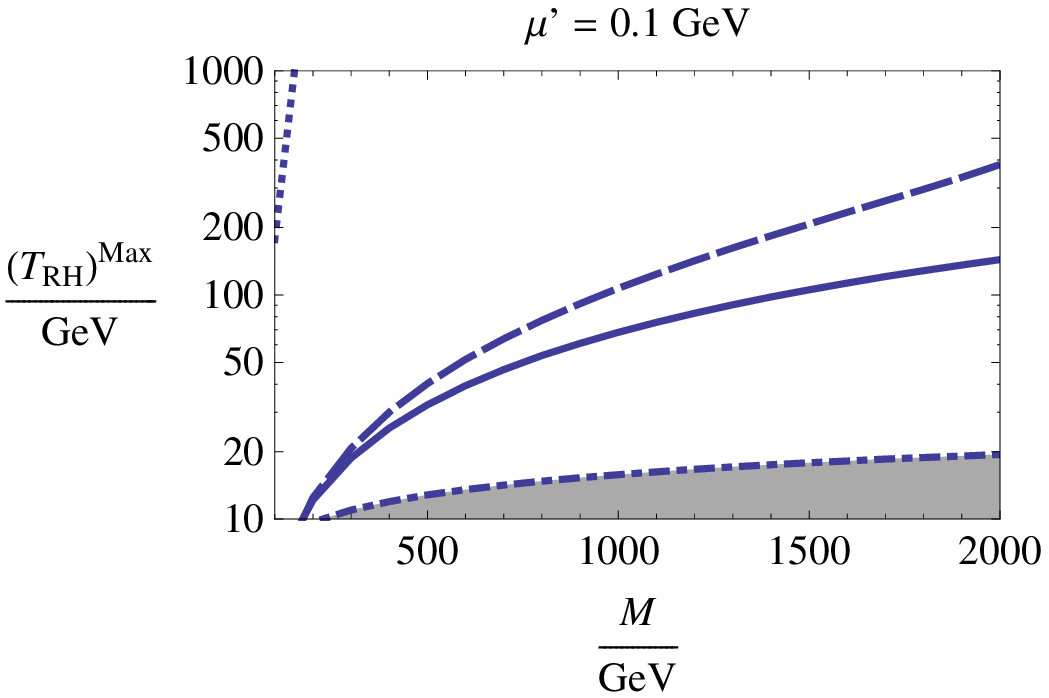} &
\includegraphics[scale=0.7]{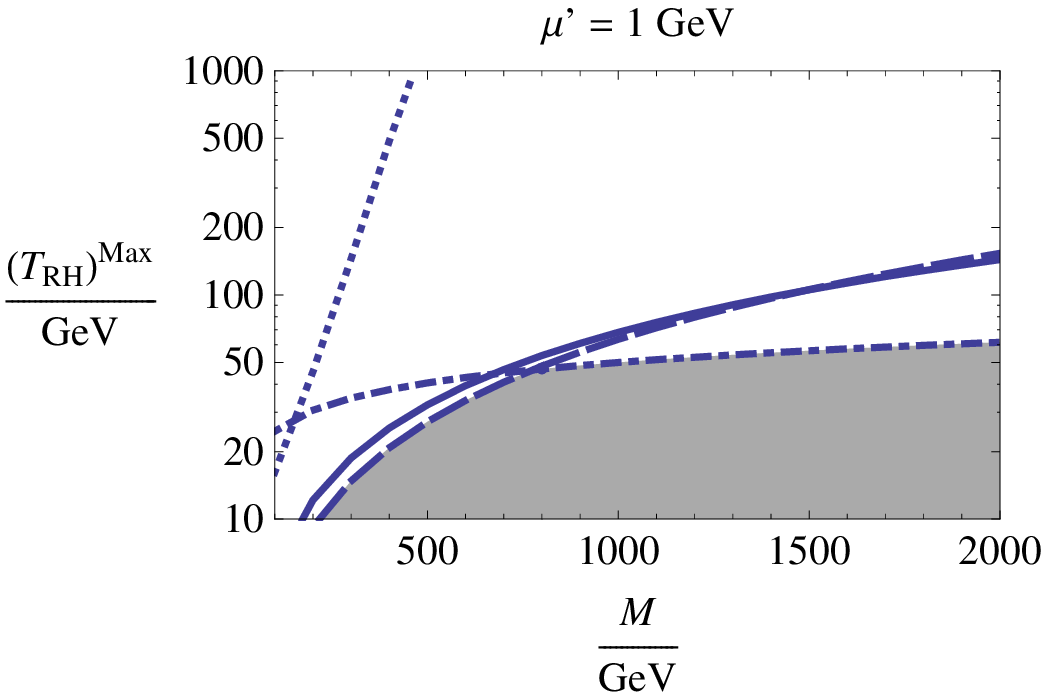}
\end{tabular}
\includegraphics[scale=0.7]{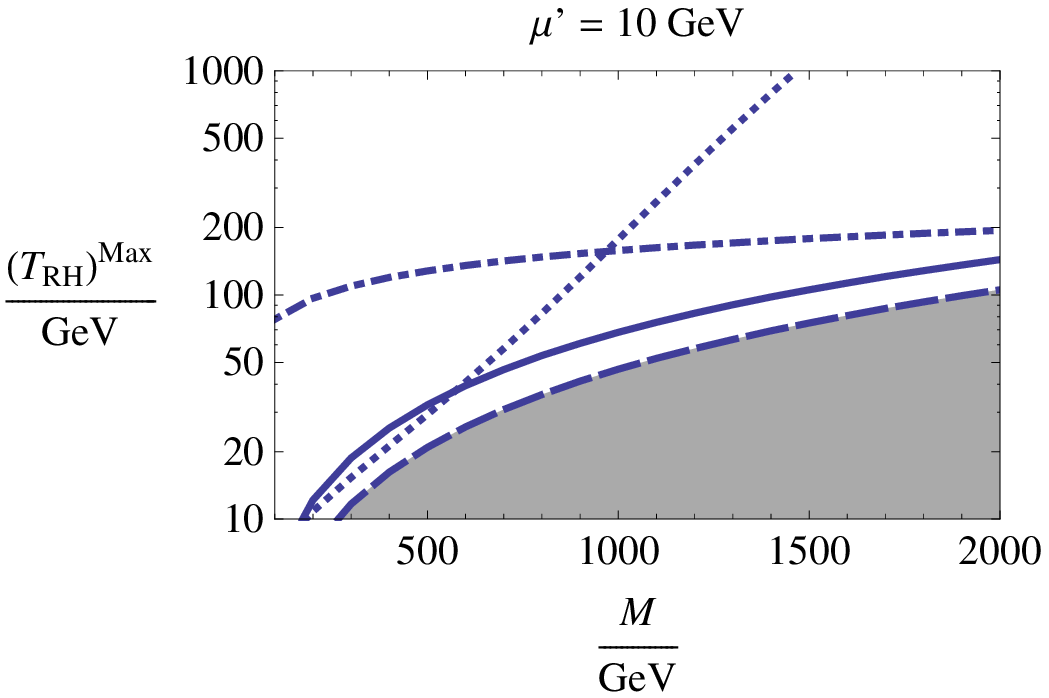}
\caption{Plots of maximum allowed value for $T_\mr{RH}$ as a function of $M$ for $\mu'= 0.1,\,1,\,10$ GeV.  The shaded region is allowed.  The curves bound the regions where $H(T_\mr{RH}) >\Gamma_\mr{washout}^\mr{heavy}(T_\mr{RH})$ (solid), $H(T_\mr{RH}) > \Gamma_\mr{washout}^\mr{light}(T_\mr{RH})$ (dotted), $H(T_\mr{RH}) > \Gamma_\mr{ID}(T_\mr{RH})$ (dashed) and $\Gamma_\mr{decay}>\Gamma_\mr{ann}(T_\mr{RH})$ (dash-dotted).  Different processes constrain the reheat temperature for the various values of $\mu'$.  We have taken $g'=0.07$ and $m_{\phi}=2\times 10^4$ GeV for the purpose of illustration.}
\label{fig:TRHMax}
\end{figure}

In Appendix \ref{sec:extimateWashoutRates} we can estimate these rates.  No washout of the baryon asymmetry occurs as long as \footnote{If these constraints hold at $T=T_\mr{RH}$ then they will hold for all subsequent temperatures since the Boltzmann suppression will always dominate over the $T^2$ dependence of $H$.}:
\bea
H(T_\mr{RH}) &>& \Gamma_\mr{ID}(T_\mr{RH}), \label{eq:IDCondition}\\
H(T_\mr{RH}) &>& \Gamma_\mr{washout}^\mr{heavy}(T_\mr{RH}), \label{eq:heavyWashoutCondition}\\
H(T_\mr{RH}) &>& \Gamma_\mr{washout}^\mr{light}(T_\mr{RH}). \label{eq:lightWashoutCondition}
\eea
Ensuring these inequalities places constraints on the four-dimensional parameter space spanned by $g'$, $\mu'$, $T_\mr{RH}$, and $m_{\phi}$.  For various values of the parameters each of the different cosmological constraints of Eqs.~(\ref{eq:DecayRatevsAnnRate}), (\ref{eq:IDCondition}), (\ref{eq:heavyWashoutCondition}) and (\ref{eq:lightWashoutCondition}) can become the most important.  As an illustration, we have shown the allowed region for the reheat temperature as a function of the LXP mass for different $\mu'$s in Fig.~\ref{fig:TRHMax}.  While not the case for the $g'$ chosen in Fig.~\ref{fig:TRHMax}, for larger $g'$, $\Gamma_\mr{washout}^\mr{heavy}$ can be the strongest constraint.

Once we have constrained $T_\mr{RH}$ and $\mu'$, we need to set appropriate values for $m_{\phi}$ and BR, which enter the expression for $Y_{D}$, Eq.~(\ref{eq:YD}).  One constraint on the inflaton mass is $m_{\phi}>2\,\tilde{M}$ so that the decays described in Sec. \ref{sec:EpochOfInflatonDecay} will be kinematically allowed.  Then the need for $Y_D\sim T_\mr{RH}/m_{\phi}$ to be large competes with keeping $\Gamma_\mr{ann}\sim T_\mr{RH}^4/m_{\phi}$ small enough to satisfy Eq. (\ref{eq:DecayRatevsAnnRate}).  We chose BR $=0.1$ as a reasonable estimate\footnote{If $\phi$ is a dilaton (i.e. if it enters the Kahler potential in the ``no-scale" form) then it will couple to other fields in the model proportional to their mass.  This can motivate large BRs.} for the branching ratio of $\phi\rightarrow\,\tilde{D_{\ell}}\,\tilde{D_{\ell}}^*$.  Note that both Eqs. (\ref{eq:YD}) and (\ref{eq:GammaAnn}) depend on BR$/ m_{\phi}$, so  a smaller branching ratio can be offset by a smaller inflaton mass.

Cosmological parameters for the three benchmark scenarios are shown in Table \ref{tab:cosmologicalParamsForBenchmark}.  Exotic squark decays can generate the BAU for a wide range of parameters.

\begin{table}[ht]
\begin{center}
\begin{tabular}{||c|c|c|c|c|c|c|c|c||}
\hline
\hline
\multicolumn{9}{|c|}{Large $\tilde{D}$ mass splittings} \\
\hline
scenario & $\frac{T_\mr{RH}}{\mr{GeV}}$ & $\frac{H}{\mr{GeV}}$ & $\frac{\Gamma_\mr{decay}}{\mr{GeV}}$ & $\frac{\Gamma_\mr{ann}}{\mr{GeV}}$ & $\frac{\Gamma_\mr{ID}}{\mr{GeV}}$ &
$\frac{\Gamma_\mr{washout}^\mr{heavy}}{\mr{GeV}}$ & $\frac{\Gamma_\mr{washout}^\mr{light}}{\mr{GeV}}$ & $\frac{m_{\phi}}{\mr{GeV}}$ \\
\hline
large splittings & 18 & $4\times 10^{-16}$ & $8\times 10^{-4}$ & $3\times 10^{-6}$ & $8\times 10^{-17}$ & $2\times 10^{-24}$ & $1\times 10^{-17}$ & $5000$\\
\hline
\hline
\multicolumn{9}{|c|}{Degenerate $\tilde{D}$ masses} \\
\hline
scenario & $\frac{T_\mr{RH}}{\mr{GeV}}$ & $\frac{H}{\mr{GeV}}$ & $\frac{\Gamma_\mr{decay}}{\mr{GeV}}$ & $\frac{\Gamma_\mr{ann}}{\mr{GeV}}$ & $\frac{\Gamma_\mr{ID}}{\mr{GeV}}$ &
$\frac{\Gamma_\mr{washout}^\mr{heavy}}{\mr{GeV}}$ & $\frac{\Gamma_\mr{washout}^\mr{light}}{\mr{GeV}}$ & $\frac{m_{\phi}}{\mr{GeV}}$\\
\hline
high $T_\mr{RH}$  & 75 & $8\times 10^{-15}$ & $4\times 10^{-5}$ & $1\times 10^{-5}$ & $1\times 10^{-15}$ & $4\times 10^{-16}$ & $6\times 10^{-20}$ & $10^5$\\
gauge mediation   & 1000 & $1\times 10^{-12}$ & $1\times 10^{-8}$ & $1\times 10^{-9}$ & $\sim 0$ & $\sim 0$ & $\sim 0$ & $10^7$\\
\hline
\hline
\end{tabular}
\caption{The numerical values for cosmological constraints described in Sec. \ref{sec:Cosmology} for the benchmark parameters of Table \ref{tab:benchmarkParams}.  All rates are evaluated at $T_\mr{RH}$.  For the given parameters, the numerical value for $\eta$ matches the WMAP5 measurement within the approximations made here.  We expect an ${\mathcal O}(1)$ uncertainty due to diagrams that we have neglected.  We have taken $\mr{BR}=0.1$, $g_*(T_\mr{RH}) = 100$, and $\tilde{m}_\mr{Z^0} = 100$ GeV.}
\label{tab:cosmologicalParamsForBenchmark}
\end{center}
\end{table}

\section{Gauge Mediation}\label{sec:gaugeMediation}
In the simplest gauge mediated models the messengers form $N$ complete $\bf{5}$ and $\overline{\bf{5}}$ representations of $SU(5)$.  This includes $N$ families of new vector-like down-type quarks, the matter content of our model.  Naively, GMSB models exhibit an exact ``messenger-parity," akin to our ``exotic-parity," which in principle could lead to undesirable long-lived relics (references \cite{Dimopoulos:1996gy, Baltz:2001rq, Jedamzik:2005ir}  that address this issue).  One motivation for models of the type presented here is to address this messenger cosmology while simultaneously generating the baryon asymmetry. In our models, the low reheat temperature avoids the thermal production of messengers.  They are instead produced in the decays of the $\phi$ field, and they subsequently undergo baryon number violating decays.  

A complete discussion of the messenger cosmology would also require a history for  $L$-type messengers, about which we remain agnostic.  The lightest one could be a dark matter candidate \cite{Dimopoulos:1996gy} or perhaps there are additional couplings which allow them to decay before the onset of big bang nucleosynthesis (BBN), see Sec. \ref{sec:darkMatter}.  Either way we assume they do not affect the BAU.  The final set of benchmark parameters in Tables \ref{tab:benchmarkParams} and \ref{tab:cosmologicalParamsForBenchmark} is appropriate for GMSB.

To implement the GMSB scenario, we replace the mass term for the exotics in the superpotential with
\be
\mathcal{W}_\mr{GMSB} \supseteq X\delta_{ij}\,D_i\,\overline{D}_j,
\ee
where $X$ is a spurion that gets a SUSY breaking vev, $\langle X \rangle  = M_X + \theta^2 F_X$.  We assume identical couplings of different generations to the $X$ field to ensure degeneracy at this order.

\subsection{New Contributions to Up-squark Masses}

The tree-level interaction between the messengers and the MSSM $u^c$ fields via the $g'$ coupling induces new contributions to the up-squark masses from $D$ loops.  These contributions could potentially spoil the flavor-diagonal nature of the gauge mediated couplings.  Typically, the leading contribution is at two-loops and is
\be
(\delta \tilde{m}_{u_R}^\mr{2-loop})^2_{ij} \approx
-\frac{1}{(16\,\pi^2)^2}\,g'_{ikm}\,g'^*_{jkm}\,g_s^2\,\frac{F_X^2}{M_X^2}.
\ee
At 1-loop, there is an accidental cancellation at $\mathcal{O}(F_X/M_X)^2$ analogous to \cite{Dine:1996xk, Chacko:2001km}.  The residual contribution is
\be
(\delta \tilde{m}_{u_R}^\mr{1-loop})^2_{ij} =
-\frac{1}{8\,\pi^2}\,g'_{ikm}\,g'^*_{jkm}\,\frac{F_X^4}{M_X^6}.
\ee
For the GMSB parameters of Table \ref{tab:benchmarkParams},  $\delta \tilde{m}_{u_R}^\mr{2-loop}\approx -65$ GeV and $\delta \tilde{m}_{u_R}^\mr{1-loop} \approx -7$ GeV, where we have taken $F_X=7\times 10^{10}\,\mr{GeV}^2$ which implies $m_\mr{SUSY} \approx 600$ GeV.  In the language of  \cite{Gabbiani:1996hi}, this leads to a flavor off-diagonal mass-insertions of size
\bea
\delta^\mr{2-loop} &\equiv& \frac{(\delta \tilde{m}^\mr{2-loop})^2}{\tilde{m}^2} \approx \frac{g'^2}{g_\mr{SM}^2},\\
\delta^\mr{1-loop} &\equiv& \frac{(\delta \tilde{m}^\mr{1-loop})^2}{\tilde{m}^2} \approx 16\,\pi^2\frac{g'^2}{g_\mr{SM}^4}\frac{F_X^2}{M_X^4},
\eea
where $g_\mr{SM}$ is the appropriate SM coupling constant.
Since these flavor violating contributions are in the up sector, the strongest constraint is $\delta \lsim \mathcal{O}(10^{-2})$ due to $D^0-\overline{D}^0$ mixing \cite{Golowich:2007ka}.  For the GMSB benchmark parameter choices, this provides a (mild) constraint on $g'$, independent of $\mu'$.

\subsection{Proton Decay}\label{sec:pdecay}
It is often stated that both baryon number violation and lepton number violation are necessary for proton decay.  This is true only if there are no non-leptonic fermions lighter than the proton.  In GMSB where a light gravitino is present, the decay  $p\rightarrow \tilde{G} + K^+$ is open.  Following \cite{Choi:1996nk} one can estimate the lifetime of the proton in these models to constrain the parameters:
\be
\mu' < 0.3\,\mr{GeV}\, \left(\frac{0.01}{g'}\right)^{1/2} \left(\frac{\tilde{m}_{\mr{SUSY}}}{600\,\mr{GeV}}\right)
\left(\frac{{m}_{\tilde{G}}}{1\,\mr{eV}}\right)^{1/2}\left(\frac{M}{10^6\,\mr{GeV}}\right).
\ee
We have imposed the proton lifetime to be greater than $2\times 10^{33}$ years  \cite{Kobayashi:2005pe} for this channel using the bound on $p \rightarrow K^{+} \bar{\nu}$.  For the GMSB parameters in Table \ref{tab:benchmarkParams}, proton decay constrains $\mu'\lsim 1.2\,\mr{GeV}$ where we have taken $m_{\tilde{G}} = 16$ eV and $\tilde{m}_{\mr{SUSY}} = 600$ GeV, corresponding to $F_X = 7\times 10^{10}\,\mr{GeV}^2$.  Future experiments could discover proton decay if this model is correct.

\section{Phenomenology}\label{sec:Pheno}
We begin this section by discussing new contributions to an assortment of low energy processes.  The need to avoid large violations of CKM unitarity will restrict $\mu'/M$.  We will find that charmed meson mixing constrains the allowed values of $g'$.  Contributions to electric dipole moments could allow a measurement of the $\mu'$ phases in upcoming experiments.  Finally, we will outline potential collider observables.  Neutron-anti-neutron oscillation bounds are not relevant in this model due to $\mu'$ suppression.

\subsection{New Contributions to the CKM Matrix}
The diagonalization procedure of Sec. \ref{sec:InteractionsInSCKMBasis} also introduces interactions with the $u_i$, $D_i$ and $W^{\pm}$.  This leads to a $3\times (3+N)$ CKM-like matrix where the interactions with the exotic squarks are suppressed by $\mu'/M$.  Using unitarity measurements on the Standard Model sector of this new matrix, one can constrain the allowed values of $\mu'/M$.  The most constraining measurement comes from \cite{Amsler:2008zzb}
\be
|V_{ud}|^2+|V_{us}|^2+|V_{ub}|^2 = 0.9999 \pm 0.001.
\ee
In our model there is an additional contribution of $N\,(\mu'/M)^2$ on the left-hand side of this equation.  Requiring this to be within 2 $\sigma$ of the measurement implies
\be
\frac{\mu'}{M} < 0.03,
\ee
where we have assumed $N=2$.

\subsection{Flavor Changing Neutral Currents}\label{sec:FCNC}
\begin{figure}[ht]
\begin{center}
\includegraphics[scale=1]{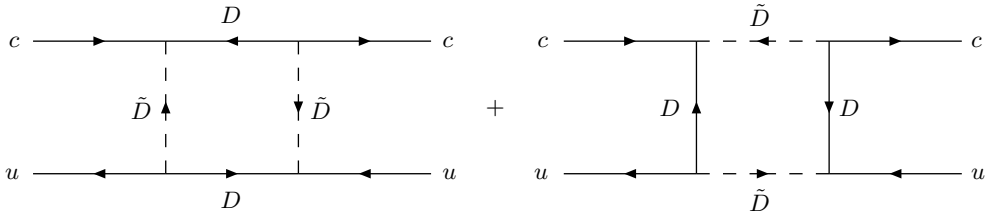}
\end{center}

\caption{New contributions to $D^0-\overline{D^0}$ mixing.}
\label{fig:DDBarMixing}
\end{figure}

Tree level interactions  between the $u$ quarks and the exotic sector with no $\mu'$ suppression (Eq.~(\ref{eq:W_Exotic})) give a potentially large contribution to $D^0$ meson mixing.  The two dominant diagrams which contribute to $\Delta M_{D^0}$ are shown in Fig. \ref{fig:DDBarMixing}.  Following \cite{Golowich:2007ka} this translates into a constraint on $g'$.  The experimental limit is $x_{D^0} < 8.7\times 10^{-3}$ with
\be
x_{D^0} \equiv \frac{\Delta M_{D^0}}{\Gamma_{D^0}} = 3.0\times 10^{-3}\,\frac{g'^4\,B_{D^0}\,f_{D^0}^2\,m_{D^0}}{M^2\,\Gamma_{D^0}}\frac{m_{D^0}^2}{(m_c+m_u)^2}
\frac{1-\left(\frac{\tilde{M}}{M}\right)^4+2\left(\frac{\tilde{M}}{M}\right)^2\mr{Log}\left[\left(\frac{\tilde{M}}{M}\right)^2\right]}
{\left[1-\left(\frac{\tilde{M}}{M}\right)^2\right]^3},
\ee
where $B_{D^0} = 0.82$, $f_{D^0} = 0.223$ GeV, $m_{D^0}$ is the $D^0$ mass; $m_u$ is the up quark mass; $m_c$ is the
charm quark mass; $\Gamma_{D^0}$ is the $D^0$ decay width; $M$ is the $D$ mass, and $\tilde{M}$ is the $\tilde{D}$ mass.
This is the source of the constraint on $g'$ in Table \ref{tab:benchmarkParams} for the second benchmark point.  Since this process only involves the first two generations, this bound can be mitigated by assuming a texture for the $g'_{ijk}$ matrix.

Contributions to neutral kaon mixing are small.  These processes are $mu'$ suppressed since they involve $d$ quarks.  The dominant contributions are from several box diagrams which are all strongly suppressed.  For the high reheat benchmark parameters (the most dangerous case) the contribution to $\Delta M_{K^0}$ is roughly 4 orders of magnitude below the observed value \cite{Amsler:2008zzb}.  Contributions to $\epsilon'_K$ from $\Delta S=1$ operators \cite{Barbieri:1985ty} which only suffer $(\mu'/M)^2$ suppression are satisfied for all the benchmarks in Table \ref{tab:benchmarkParams}.  

\subsection{Electric Dipole Moments}
In principle, there are contributions to electric dipole moments (EDM) due to the phase responsible for the BAU.  Naively, an $\mathcal{O}(1)$ phase at the TeV scale is dangerous.  However, in this model, the contributions to EDMs are suppressed by powers of $\mu'/M$.  If the exotics are the messengers of GMSB, the smallness of this ratio suppresses EDMs beyond anything that would be measured.  On the other hand, if the exotics are at the TeV scale,  the EDMs are rendered small enough to avoid current bounds but could be generated at an interesting level.

\begin{figure}[ht]
\begin{center}
\includegraphics[scale=1]{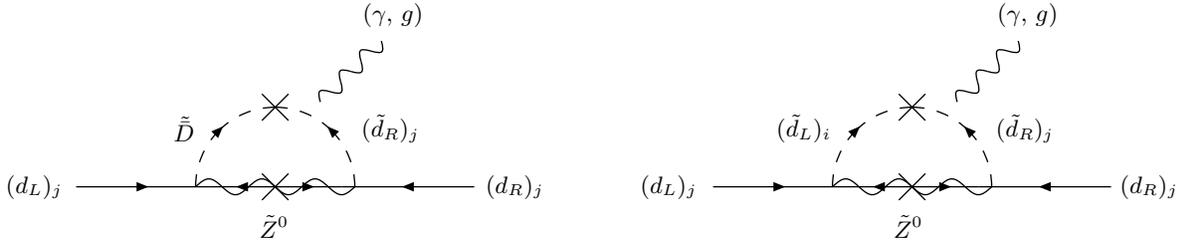}
\end{center}
\caption{Examples of 1-loop contributions to EDMs.  The squark mass insertions are $b_{\mu'_R}$ and $\delta^{LR}_{ij}$, $i \neq j$, for the first and second diagrams respectively.}
\label{fig:EDMs}
\end{figure}

To generate a non-vanishing EDM requires two elements beyond CP violation:  flavor mixing and left-right mixing.  There are two types of contributions (see Fig. \ref{fig:EDMs}).  We consider each in turn.

First, we examine the left diagram of Fig. \ref{fig:EDMs}.  The left vertex uses the interaction of Eq. (\ref{eq:DdZCoupling}).  
To close the diagram requires the analytic SUSY breaking insertion $b_{\mu'_R}\,\tilde{d_R}\,\tilde{\bar{D}}$.  At present, the most stringent constraints on hadronic EDMs come from the bound on the EDM in mercury \cite{HgEDM}, $d_\mr{Hg} < 3.1 \times 10^{-29}$ $e$-cm.  Following  \cite{HgTheory}, we estimate 
\be\label{eq:EDMwithbmu}
d_\mr{Hg} \approx 2.6 \times 10^{-29}\,e\,\mr{cm}\,\left(\frac{\mu'_{L}/M}{10^{-3}}\right)\,\left(\frac{b_{\mu'_R}/M^2}{10^{-2}}\right)\left(\frac{500\,\mr{GeV}}{M}\right)^{2}\left(\frac{\tilde{m}_{Z^{0}}}{100 \, \mr{GeV}}\right),
\ee
Requiring the EDM to be below the current bound effectively places a constraint on $b_{\mu'_R}$.  For the first benchmark point, there are no other requirements on the size of this coupling\footnote{For $b_{\mu'}\sim \mu'\,M$ there are modifications of $\epsilon$ comparable to the ones calculated here.}.  So, this class of diagrams could contribute to an EDM very close in size to the current bound.  We have employed the QCD sum rules approach to calculating the EDM.  In this case, the relevant $\mu'$ corresponds to mixing with the down quark. This is constrained by considerations of fine-tuning, see Eq.~(\ref{eq:finetuning}).  
For cases where degeneracy is important in generating the BAU, $b_{\mu'}$ is constrained by the requirement that enforces the degeneracy of the squarks (see Appendix \ref{sec:SmallSplit} for a detailed discussion).  The EDM arising from this class of diagram will be small when a resonant enhancement of $\epsilon$ is required.

There is another diagram that can give contributions of an interesting size. For the second diagram in Fig. \ref{fig:EDMs}, the vertex on the left comes from an interaction induced by the rotations to eliminate  $\mu'_L$ described in Sec. \ref{sec:InteractionsInSCKMBasis}.  To see the importance of this rotation, note that in the SUSY limit there is a correction to the $d_L-d_L-Z^0$ coupling at $\mathcal{O}(\mu'^2)$.  While hermiticity of the Lagrangian ensures the reality of diagonal interactions with gauge bosons, new off-diagonal $(d_L)_i-(d_L)_j-Z^0$ interactions can inherit CP violation from $\mu'_{L}$.  It is the supersymmetric analog of this interaction that appears in the left vertex.  To couple to $d_R$ and close the loop then requires an analytic off-diagonal insertion.  This leads to an EDM for Mercury of the size
\be
d_\mr{Hg} \approx 3 \times 10^{-33}\,e\,\mr{cm}\,\left(\frac{\mu'_{L}/M}{10^{-3}}\right)^2\,\left(\frac{\delta^{LR}_{1i}}{10^{-3}}\right)\,\left(\frac{500\,\mr{GeV}}{m_\mr{SUSY}}\right)^2 \left(\frac{\tilde{m}_{Z^{0}}}{100 \, \mr{GeV}}\right),
\ee
where $\delta^{LR}_{1i}$ is the flavor off-diagonal mass insertion.  While this contribution appears small, this may be a artifact of the QCD sum rules approach used in the calculation of the mercury EDM.  Parton quark model \cite{PQM} calculations of the neutron EDM indicate that the strange quark can yield a large contribution.  Assuming that this also holds for the Hg nucleus, we can make the replacement $\delta_{1i} \rightarrow \delta_{23}$, which is only bounded by measurements of the branching ratio for $b \rightarrow s\, \gamma$, implying a constraint $\delta_{23}^{LR} \lsim 10^{-2}$.  Also, a $\mu'_{L}/M$ as large as  $10^{-2}$ for the strange quark is allowed without fine-tuning the strange quark mass.  All told, a Hg EDM as large as ${\mathcal O}(10^{-30}) \, e$-cm might be produced by this class of diagram, which might be visible in future experiments.  Experiments searching for EDMs of the deuteron or neutron might also be sensitive to the phases in $\mu'$.

Depending on the size of the supersymmetry breaking parameters, it may be possible to observe the CP violation responsible for generating the BAU.  However, the BAU does not depend on supersymmetry breaking, so it is possible that the EDMs might be unobservable without affecting the size of the generated asymmetry.

\subsection{Dark Matter}\label{sec:darkMatter}
Due to the R-parity violating interactions in Eq. (\ref{eq:uddCoupling}), the LSP can decay.  For neutralinos the decay channel will be $\tilde{\chi}^0\rightarrow\,u\,d\,d$ with a rate
\be
\Gamma_{\tilde{\chi}^0} \approx \frac{g'^2 g^2_{\tilde{\chi} d \tilde{d}}}{64\,\pi^3}\left(\frac{\mu'}{M}\right)^4\,\frac{\tilde{m}_{\chi^0}^5}{\tilde{m}_d^4},
\ee
where $g_{\tilde{\chi} d \tilde{d}}$ is the coupling between the neutralino, the down-type quark, and down-type squark; $m_{\chi^{0}}$ is the neutralino mass, and $\tilde{m}_d$ is a down-type squark mass.   In the large splittings (i) and high reheat (ii) scenarios the lifetimes are $\mathcal{O}(10^{-11}\,\mr{s})$ and $\mathcal{O}(10^{-5}\,\mr{s})$ respectively -- short enough to decay before BBN.  In general, the requirement that the LSP decay before BBN places a weak lower bound on $\mu'/M$.  For the gauge mediated benchmark (iii), the gravitino is the LSP.  In this case, the relevant cosmological constraints are for the NLSP, which decays via $1/F$ suppressed couplings (see \cite{Giudice:1998bp} for a discussion).  Note that even though there is R-parity violation, the gravitino is stable since it is lighter than the proton and lepton number is conserved.

Since the LSP decays for benchmarks (i) and (ii), it is no longer a viable dark matter (DM) candidate.  An additional stable weak scale particle must be present.  Note, the reheating temperature is sufficiently high that this particle (whatever its identity) may have a thermal history.  If the exotic quarks are embedded into a $5$ and $\overline{5}$ representation of $SU(5)$, the lepton doublets ($L$, $\overline{L}$) could fulfill this role.  However, the simplest DM models with these quantum numbers are ruled out due to coherent scattering off nuclei in direct detection experiments.

\subsection{Collider Signatures}\label{sec:colliderSigs}
There are a number of LHC signatures whose observation would lend support to this mechanism as the source of the BAU.  After confirming SUSY, one would need to determine the existence of the exotic quarks or squarks.   Exotic heavy colored states at the LHC without baryon number violating interactions have been investigated in \cite{Kang:2007ib, Gerbush:2007fe, Arik:2001bc, DelNobile:2009st}, so we will not discuss issues related to discovery any further.  Instead, we concentrate on phenomenology induced by the baryon-violating couplings.

Much of the novel phenomenology actually arises in the MSSM sector, and does not rely on the direct production of the exotics.  In particular, the decay of the LSP opens several new possibilities for collider phenomenology. LSPs that would otherwise be constrained (squarks and charged leptons) are now possible.  All MSSM LSP decays will be suppressed by $(\mu'/M)^4$, and over much of the parameter space can lead to displaced vertices.

Whether or not the LSP decay gives an observable displaced vertex depends sensitively on the identity of the LSP.
If MSSM squarks are the LSP, they decay directly to two quarks with path length
\be\label{eq:squarkDecayLength}
c\, \tau_{\tilde{q}} \approx 1 \textrm{ mm} \times \left( \frac{0.1}{g'} \right)^{2}  \left( \frac{10^{-3}}{(\mu'/M)} \right)^{4} \left( \frac{1 \textrm{ TeV}}{\tilde{m}_q} \right).
\ee
If the charginos or neutralinos are the LSP, they can decay to 3 jets with path length
\be
c\, \tau_{\tilde{\chi}} \approx 40 \textrm{ cm} \times \left( \frac{0.1}{g'} \right)^{2}  \left( \frac{1.0}{g_{\tilde{\chi} q \tilde{q}}} \right)^{2} \left( \frac{10^{-2}}{(\mu'/M)} \right)^{4} \left( \frac{\tilde{m}_q}{1 \textrm{ TeV}} \right)^{4} \left( \frac{100 \textrm{ GeV}}{m_{\tilde{\chi}}} \right)^{5}.
\ee
Here, $g_{\tilde{\chi} q \tilde{q}}$ is the neutralino-quark-squark coupling.  Slepton and sneutrino LSP decays are at least four-body (to conserve lepton number), and  such decays are too slow to be observed on detector time scales.  If instead the gravitino is the LSP, then decays of the NLSP to the gravitino are much faster than R-parity violating decays.  In this case collider signatures will be the same as in ordinary gauge mediation.

The best way to substantiate this baryogenesis mechanism is to observe baryon number violating decays.
There are four different possibilities which are favorable for the direct observation of $B$-violating decays:
\begin{enumerate}
\item 2-body $B$-violating decays of LXPs,
\item 2-body $B$-violating decays of LSP squarks with (or without) displaced vertices,
\item 3-body $B$-violating decays of LSP charginos with long-lived tracks,
\item 3-body $B$-violating decays of LSP neutralinos with displaced vertices.
\end{enumerate}

To be certain a given decay is $B$-violating one must establish the baryon number of the parent particle.  Alternately, this requires observing two different processes:  one which preserves $B$ (or alternately, defines the baryon number of the parent particle) and one which violates it.  In the case of the LSP (2-4 above), there is only a single type  of decay (baryon number violating).  The baryon number of the LSP must be established by a means other than decay.  One method is examining other particles present in the cascade down to the LSP.  A second is by searching for associated production processes of the LSP.  For the case of exotic squark decays (1), one could imagine measuring both $\tilde{D}\rightarrow \tilde{\chi}^0 + d^{\dagger}$ and $\tilde{D} \rightarrow u+d$ which would be definitive proof of $B$-violation.

Finally, we point out decays of the type $X \rightarrow q\, q$ are exotic in their own right.  The two quark final state is indicative of either baryon violation (as here, where $X$ can either be the exotic $D$ or a squark LSP) or the existence of an exotic diquark with baryon number two. We now comment on the possibility of observing these decays.  Ideally, we would like to verify that the final state is quarks, and not a quark-anti-quark pair.  This is most easily accomplished if the final state consists of a top and bottom quark.  The charge of the top quark can be determined by the charge of the $W^{\pm}$ via its leptonic decays.  Then, semi-leptonic $b$-quark decays may be used to determine the charge of the initial $b$-quark.  Competition between semi-leptonic decays of the bottom quark, $b\rightarrow \ell\,+...$, and hadronic decays of the $b$-quark followed by semi-leptonic decays of the charm quark, $b\rightarrow \overline{c}\,+... \rightarrow \overline{\ell}\, +...$, dilute measurements of the bottom charge.  To get some idea of how hard it will be to observe $B$-violation, we will do a simple estimate of how many events should have the right final states to reconstruct the 2-body LXP or LSP squark decays to $t+b$ at the LHC.  In the case of the LSP, it is quite likely that there will be a displaced vertex, which should be useful in eliminating any SM background.  The true challenges are the combinatorics for reconstructing a complicated final state (e.g., $t\,b\, \bar{t}\, \bar{b}$), the overall event rates, and the relevant tagging efficiencies.

At $\sqrt{s} = 14$ TeV, the squark-squark production cross section (for $m_{\tilde{g}} =$ 1 TeV) is $\sim 1$ pb for $m_{\tilde{q}} = 800$ GeV, and $\sim 5$ pb for $m_{\tilde{q}}=600$ GeV \cite{Beenakker:1996ch}, so we will have $\mathcal{O}(10^5)$ squark pairs produced in 100 $\mr{fb}^{-1}$ of data.  We assume a 100\% branching ratio to the $B$-violating channel $t+b$, which is equivalent to imposing a large hierarchy in the $g'$ couplings. We assume a 40\% efficiency for $b$-tagging.  Using the $b$-quark semi-leptonic branching fraction of 20\% and the $W^{\pm}$ leptonic branching fraction of 22\% (electron and muon inclusive), we find $10^{3}$ baryon number violating squark decays will be $b$-tagged with the appropriate leptonic final states. Whether or not this sample will be sufficient to determine the sign of the decaying particles is a detailed experimental question. It does not seem unreasonable that such a measurement would be possible.

Because the presence of a (nearly) degenerate pair of squarks is required for the generation of a sufficient asymmetry over much of the parameter space, it would be important to verify the presence of this duplicity of states.  It appears impossible to resolve the squarks at the LHC via simple mass measurements. However, a measurement of the production cross section might indicate the presence of the second degenerate squark.

\section{Conclusions and Future Work}
The baryon number violating decay of exotic squarks can generate the BAU for a wide range of parameters.  This scenario is constrained by both cosmological arguments and particle experiments.  The observation of exotic squarks at the LHC would be the first step towards verification of this scenario.  The subsequent observation of baryon number violating decays would be a smoking gun.  Alternately, even if no exotics are observed, if the MSSM spectrum is consistent with gauge mediation, the observation of proton decay would be supportive of this mechanism.

The goal of this work is to argue the viability of a simple idea: exotic squarks generated the BAU.  Detailed exploration of parameter space is left for future work.  In addition, different assumptions about the pattern of supersymmetry breaking might induce additional contributions to the BAU.   Also, for some regions of parameter space, the exotic quarks could be the LXP.   In this case, the generation of the BAU proceeds in a nearly identical fashion.  

When very small mass splittings are required, one would like to see specific examples of the family symmetries discussed in Appendix \ref{sec:SmallSplit} to realize the squark degeneracy.  This may lead to correlations between a solution to the SUSY flavor problem and the BAU.  

Due to R-parity violation, the LSP is no longer a viable dark matter candidate.  It would be interesting to explore models which could extend the exotic sector to include a new dark matter state, thereby connecting the BAU and dark matter.

Should exotic squarks, their baryon number violating decays, and non-vanishing electric dipole moments all be observed, we will have gone a long way towards establishing the origin of the baryon asymmetry of our universe.

\vspace{-0.2cm}
\subsection*{Acknowledgments}
\vspace{-0.3cm}
\noindent
We thank Dan Amidei, Kenji Kadota, Eric Kuflik, Markus Luty and Jesse Thaler for useful discussions.
The work of T.C. was supported in part by the NSF CAREER Grant NSF-PHY-0743315.
The work of  D.J.P. was supported  in part by DOE Grant \#DE-FG02-95ER40899 and by a Rackham Pre-doctoral Fellowship.
The work of A.P. was supported in part by NSF Career Grant NSF-PHY-0743315 and by DOE Grant \#DE-FG02-95ER40899.  A.P. also acknowledges the Aspen Center for Physics, for providing a stimulating environment where some of this work was completed.

\def\theequation{\Alph{section}.\arabic{equation}}
\begin{appendix}

\setcounter{equation}{0}
\section{Engineering Small LXP Splittings}\label{sec:SmallSplit}
We assume exotic-parity is a remnant of an unspecified family symmetry which acts on the exotic sector SUSY and soft-breaking parameters.   The $\mu'$ coupling explicitly breaks this family symmetry along with exotic-parity.  In the absence of the $\mu'$, the squarks would be exactly degenerate -- all splittings are proportional to $\mu'$.

Specifically, we assume the symmetry enforces
\bea
M_{ij} &=& M\,\delta_{ij},   \label{eq:Mdegen} \\
\tilde{M}_{ij} &=& \tilde{M} \,\delta_{ij},  \label{eq:MSqdegen} \\
(b_M)_{ij} &=& b_M\,\delta_{ij}. \label{eq:bMdegen}
\eea
This will keep new CP violating phases from being generated in the exotic sector.  Since $\mu'$ is assumed to be the only parameter which breaks the symmetry, all soft-masses which mix the exotics with the MSSM will be proportional to this parameter. 

How large a splitting do we expect between the squarks?
We first assume Eqs. (\ref{eq:Mdegen})--(\ref{eq:bMdegen}) hold.  The introduction of a non-zero $\mu'$ induces $\Delta \tilde{M}^2 \sim \mu'^2$.  There are additional potential sources of splitting.  As an example, assume that the $b_{\mu'}$ term is non-zero.    If $M\sim m_\mr{SUSY}$ the mass splittings are proportional to $b_{\mu'}$.  If instead there is a large hierarchy $M \gg m_\mr{SUSY}$, e.g. for the GMSB scenario, then the splittings are proportional to $b_{\mu'}^2/M^2$.  Care must be taken to make sure these  splittings induced by $b_{\mu'}$ do not upset the resonance condition, i.e. $\Delta \tilde{M}^{2} \sim \mu'^{2}$.  This implies $b_{\mu'} \sim \mu'^2$ when $M \sim m_\mr{SUSY}$, while for larger $M$ (GMSB scenario), the weaker condition $b_{\mu'}  \sim  \mu' \, M$ is required.

To summarize, in order to achieve degenerate squarks we assume the mass terms obey the following properties:
\begin{itemize}
\item The superpotential mass is proportional to the identity: $M_{ij} = M\,\delta_{ij}$.
\item The non-analytic and analytic exotic squark mass matrices are proportional to the identity:
$\tilde{M}_{ij} = \tilde{M}\,\delta_{ij}$ and $(b_M)_{ij} = b_M\,\delta_{ij}$.
\item The mixing between the MSSM and the exotic sectors ($\tilde{d_R}\,\tilde{D}$, $\tilde{d_L} \tilde{D}^*$, etc.) is constrained.  While the precise size of the mixing depends on the size of $M$, it will at minimum need to be proportional to $\mu' \, M$.
\item The MSSM soft-terms are left unspecified.
\end{itemize}

Now that we understand the conditions at tree-level we will discuss the radiative stability of these requirements.  Non-renormalization of the superpotential ensures that we  need only worry about contributions to wave function renormalization.  If the $g'$ coupling breaks the family symmetry, loop effects will introduce scalar mass squared splittings of
$\mathcal{O}(\frac{g'^2}{16\,\pi^2} M^{2})$ which can potentially be larger than $\mu'^{2}$.  We need to make sure that this contribution is either sub-dominant or proportional to the identity matrix in flavor space.  The latter can be done by imposing specific textures on the $g'$ coupling matrix.  The form of these textures depends on the number of exotic families.  The flavor structure of the wave function renormalization at 1-loop comes from
\be\label{eq:WavefunctionRenormCoef}
Z^{D}_{ij} \sim g'_{ki\ell}\,(g'_{kj\ell})^*
\ee
where we are neglecting contributions $\mathcal{O}(\mu'/M)$ since these will not generate
splittings larger then the tree-level contributions.  For $N=2$, the
couplings have the form
\be
g'_{ijk} = \left( \begin{array}{cc}
              0 & g'_i  \\
             -g'_i  & 0
\end{array} \right) = g'_i\,\epsilon_{jk}.
\ee
Then it is easy to see that Eq. (\ref{eq:WavefunctionRenormCoef}) becomes
\be
g'_{ki\ell}\,(g'_{kj\ell})^* = (g'_k)^2\,(\epsilon_{i\ell}\,\epsilon_{j\ell}) = -2\,(g'_k)^2\delta_{ij}. 
\ee
In the case where $N=2$, once we achieve splittings of $\mathcal{O}(\mu'^2)$
they will be radiatively stable.  For $N>2$, specific textures on the $g'$ couplings will be required for degeneracy to hold at loop level. 

\section{Estimates of Washout Rates}\label{sec:extimateWashoutRates}
In this appendix we will estimate the rates for the three types of washout processes relevant for this model.  We will also discuss sphaleron processes which are relevant for scenarios where the squarks are the messengers of GMSB.

\subsection{Inverse Decays}
For inverse decays in Eq. (\ref{eq:inverseDecay}) the relevant rate is given by \cite{Kolb:1979qa}
\be
\Gamma_\mr{ID} = Y_D^\mr{eq}\,\Gamma_\mr{decay}^{B\mr{-violating}},
\ee
where $Y_D^\mr{eq}$ comes from applying the momentum conserving delta function to the distribution functions for $u$ and $d$ and integrating over phase space, and $\Gamma_\mr{decay}^{B\mr{-violating}}$ is given by the first diagram in Fig. \ref{fig:GammaTotalD}.  
Using the same level of approximation described in Sec. \ref{sec:BaryoFromDecays} to calculate the $B$-violating width gives
\be
\Gamma_\mr{ID}(T_\mr{RH}) \approx \frac{45}{4\sqrt{2}\,\pi^{7/2}\,g_*}\left(\frac{M}{T_\mr{RH}}\right)^{3/2}\,e^{-M/T_\mr{RH}}\,\left(\frac{9\,g'^2\,\mu'^2}{32\,\pi\,M}\right).
\ee

\subsection{Light Final States}
To determine the rate for the process in Eq. (\ref{eq:washout1}), we need the number density of the $\tilde{Z}^0$s.  LSPs are produced in the cascade decays of all exotics and MSSM superpartners which were created by the decays of $\phi$. If either the annihilation rate or decay rate of the LSPs is sufficiently fast, the  LSPs ``instantaneously" re-thermalize and follow a non-relativistic equilibrium Boltzman distribution.  Otherwise, one should use the $n_{\tilde{Z}^0}$ from non-thermal production.  Which of these number densities is appropriate depends on $T_\mr{RH}$ (in the case of annihilation) or the size of the coupling in Eq. (\ref{eq:uddCoupling}) (in the case of decay).  For the benchmark models, the LSPs do indeed re-thermalize, so the washout rate is given by:
\bea
\Gamma_\mr{washout}^\mr{light}(T_\mr{RH}) &=& n_{\tilde{Z}^0}\,\langle \sigma_\mr{washout}^\mr{light}\,v\rangle\\
                                 &\approx& 12 \left(\frac{\tilde{m}_{Z^0}\, T_\mr{RH}}{2\,\pi}\right)^{3/2}\,e^{-\frac{\tilde{m}_{Z^0}}{T_\mr{RH}}} \, \left(\frac{2\,g'\,g_w}{4\,\pi}\right)^2\,\left(\frac{\mu'}{M}\right)^4\,\frac{T_\mr{RH}^2}{(m_\mr{SUSY}^2+T_\mr{RH}^2)^2}
\eea
where $n_{\tilde{Z}^0}$ is the non-relativistic number density of Zinos, and the label ``light" refers to the masses of the final states.  We have used a simple estimate for the thermally averaged cross section.  The factor of 12 accounts for the contributions from all MSSM quarks and squarks. We have assumed that the bottom quark is ``light" for the reheat temperatures considered here.  For the GMSB benchmark where $T_\mr{RH} \gg m_{\tilde{Z}^0}$, the Zino will be relativistic so that $n_{\tilde{Z}^0}\sim T_\mr{RH}^3$.  Since $\langle \sigma_\mr{washout}^\mr{light}\,v\rangle \sim (\mu'/M)^4$, the large $\mu'$ suppression keeps this rate negligible for the GMSB scenario.

\subsection{Heavy Final States}
For the process in Eq. (\ref{eq:washout2}) more care is required.  The reason is two-fold:  there will be Boltzmann suppression since only the tails of the $u$ and $g$ distributions have enough energy to create the exotic pair, and the initial states are both relativistic at the time of
freeze-out so Maxwell-Boltzmann statistics do not apply for the normalization.  Modifying the results of \cite{Gondolo:1990dk} we find:
\bea
   \langle \sigma_\mr{washout}^\mr{heavy}\, v\rangle &=& \frac{2\,\pi^2\,T}
   {\left(\int_0^{\infty}4\,\pi\frac{E^2}{\mr{Exp}(E/T)-1}\mr{d}E \right)
   \left(\int_{m_u}^{\infty}4\,\pi\frac{E\sqrt{E^2-m_u^2}}{\mr{Exp}(E/T)+1}\mr{d}E \right)}
   \times \nonumber \\
   &&\int_{4\,M^2}^{\infty}(\sigma_\mr{washout}^\mr{heavy})(s-4\,m_u^2)\sqrt{s}\,K_1(\sqrt{s}/T)\mr{d}s,
\eea
where $K_1$ is the modified Bessel function and the label ``heavy" refers to the mass of the final states.
The factor of $K_1$ assumes Maxwell-Boltzmann statistics which is a
good approximation since the integral is evaluated from $s=4\,M^2$ to $\infty$, which is larger than $T_\mr{RH}$ in this study.

To obtain a rate we multiply this thermally averaged cross section by the relativistic number density for the gluons:
\be
n_g = 1.2\,\pi^2\,(2\times 8)\,T^3,
\ee
to get
\be
\Gamma_\mr{washout}^\mr{heavy} = n_g\,\langle \sigma_\mr{washout}^\mr{heavy}\, v\rangle.
\ee

In order to evaluate this numerically we approximate the total washout cross section by the dominant process which proceeds via a t-channel $D$ quark (see Fig. \ref{fig:WashoutProcesses}).
 In Fig. \ref{fig:TRHMax} we show the maximum allowed reheat temperature as a function of the $M$ for all the washout rates.

\subsection{Sphaleron Processes}
For the GMSB benchmark, $T_\mr{RH}> m_W$ which implies that electroweak sphaleron rates are unsuppressed.  Our mechanism for generating the baryon asymmetry does not generate any associated lepton asymmetry.  This implies that we have a non-zero value for $B$ but not $L$ (where $L$ is the lepton number of the universe).  Since sphalerons only violate $B+L$, all they will do is distribute the $B$ generated from the exotic squark decays to both $B$ and $L$.  The result is that the sphalerons will reduce $\eta$ by approximately a factor of 2.  For the exact relationship between the initial $B-L$ and final $B$ and $L$ asymmetries, see \cite{Harvey:1990qw}.
\end{appendix}

\bibliography{ESB}

\end{document}